\documentstyle[12pt]{article}
\begin{document}

\rightline{IMSc/97/06/20}

\vskip 0.5in

\centerline {\large {\bf HAVE WE AT LAST FOUND THE A-TOMS?}}

\vskip 1in

\begin{center}
{\bf R. Ramachandran} \\
The Institute of Mathematical Sciences \\
Taramani, Chennai (Madras) 600 113 \\
\end{center}

\vskip 1.5in

\begin{abstract}
It is argued that the three families of quarks and leptons are the
building blocks of all matter and all forces  (leaving out gravity) among them
are mediated by photons ($\gamma$), the weak bosons ($W^\pm$ \ and \ $Z$) and
the gluons.
\end{abstract}

\newpage

\noindent \underline{{\bf INTRODUCTION}}

\vskip 0.3in

From time immemorial, it has been the human endeavour to know what
everything may be made up of and what  is the form of all forces of
nature.  Can we understand a complex system -- say the universe -- in
terms of its parts, which in turn is made up of further parts?  Is there
a basic set of constituents in terms of which, in principle, all matter
and all phenomena of the physical world are understandable?  Is it possible
to reduce them to just a few principles that can be splashed on a
T-Shirt?

\medskip

Hindu texts refer to `Pancha Bhutas': everything is a manifestation of
a harmonious combination of Prithvi (earth), Jala (water), Vayu (air),
Agni (fire) and Akasha (space?).  Greek philosophers also mention four
of these elements, leaving out Akasha to begin with and later realise
that they need a `void' to place the other elements in.  While today we
may recognise these not as basic constituents, but merely as different
physical attributes of all matter, it would be nice if  there are just a
few ingredients from which everything can be built.
 It was Democritus of Mellitus, in
Greece, who reasoned that as we keep dividing anything into smaller and
yet smaller pieces, the process may terminate, yielding at the end
ultimate constituents, that may not be cuttable any further. He named
them a-tom (tom in Greek is to cut; a-tom means that which is
uncuttable).  Unfortunately the then dominant school of philosophy led by
Aristotle largely ignored, in fact rediculed, Democritus and his notion
of ultimate constituents.

\medskip

Several centuries later, we have the English chemist Dalton, who
resurrected the idea of basic elements of nature and propounded that the
large variety of chemical compounds can be understood in terms of
just about 90 elements -- each element being made up of its basic unit, which
he called the atom, making use of the terminology of Democritus.  
All chemical compounds are molecules -- made up of the atoms of the
elements it is  composed of.  These atoms were indeed uncuttable in
the length scales of chemistry (a few nanometres ($10^{-9}m$)), but now
we are aware that the atoms in Bohr model consist of a much tinier 
($10^{-15}m$)
massive nucleus and very light electrons moving around the nucleus,
somewhat like a miniature solar system.  The nucleus has protons and neutrons as
its constituents and they too reveal substructure.  Indeed the name atom
for these species was rather premature.  They are certainly not the
a-toms of Democritus. 

\medskip

So let us carry on the quest for the real a-toms of Democritus.  Dalton's 
atoms are
made up of protons and neutrons in the nuclei and electrons which are 
bound to them by Coulomb forces.  All chemical reactions are complex
manifestation of the electromagnetic forces experienced by the
electrons, so indeed is 
the short range van der Waals force between atoms and molecules
-- a consequence of the polarisability of the neutral atoms and
molecules. The forces of
elasticity of bulk material can be traced to the same electromagnetic 
origin.
Indeed most of the chemical and physical phenomena are understandable
in terms of the forces of electrodynamics, the classical content of
which is summarised by the set of four equations of Maxwell.  Let us probe further to
see what the protons are made up of.  What is the force that holds the
protons and neutrons together in the nucleus, overcoming the Coulomb
repulsion among the protons?  What are the analogues of Maxwell's
equation for the strong nuclear forces?  What is the spectrum of states
of which protons and neutrons are members.

\medskip

In the sixties with the advent of proton and electron accelerators,
there was a discovery of a plethora of states, which were either excited  
nucleons and similar states  on the one hand or many mesons that
could be playing the role of the strong nuclear force mediators on the 
other. Pion
or $\pi$-meson was predicted by Yukawa as the particle that could be
principally responsible for the attractive short range force between
nucleons. Pions were seen during 50's in cosmic ray experiments and were 
later produced copiously in the particle accelerators at Rochester, 
Columbia and Chicago. A prominent resonance in the pion-nucleon 
scattering was
observed at a total energy of 1238 MeV and understood as a shortlived state 
$\Delta$,  the lifetime being so
small that it shows up as a resonance with a width of 120 MeV. More resonances were discovered and many new mesons 
were produced in the accelerator laboratories.  Soon there was a flurry of 
activity in Particle Physics,
classifying and jotting down the properties of the various species of
particles and the many short lived and long lived states were systemetised.  
The particle
spectroscopy was studied, naturally  with a view to finding the clues for the
underlying dynamics.  In this way we discovered  
several new quantum numbers such as strangeness
$S$ and charm $C$, that were preserved in the strongly interacting
production processes, but violated in its  weaker decays.  The
weak forces were the ones responsible for  radioactivity in some (rather,
many) nuclei.  Particle Physics is the discipline that systematises the
subconstituents of the subatomic world and strives to answer whether the
a-toms of Democritus  can indeed be real and arrived at.

\vskip 0.3in

\noindent \underline{{\bf STANDARD MODEL}}

\vskip 0.2in

When the Russian chemist Mendeleev gave the periodic table of elements,
he  paved the way for the systematic understanding of the atomic
structure of elements.  Further during the first half of this 
century,  the
development of quantum mechanics yielded a fairly complete qualitative and
quantitative confirmation of the underlying dynamics of all chemical
phenomena.

\medskip

In the world of Particle Physics we are now in a position of having got
the analogue of the periodic table that classifies the ingredients of all
matter and this clearly points towards the nature of the underlying forces.  
This is referred to as the {\it Standard Model}.  The remarkable
feature of the Model, which we claim  explains all forces of
interaction (except gravity) among all species of matter, is that the laws are
simply suitable generalisation of the familiar Maxwell's equations that
govern all electromagnetic phenomena.

\medskip

The main principle in the Standard Model is to elevate the notion of
symmetry to the status of the essence of dynamics.  Recall that the Maxwell's
Equations are:

\medskip

$$
\begin{array}{rclccrcl}
 \bf{\nabla} . \bf{E} & = & \displaystyle{
{\rho \over \varepsilon_0}} &; &
 \bf{\nabla} . \bf{B} & = & 0 \\
 \bf{\nabla} \times \bf{E} + 
 \displaystyle{{\partial \bf{B} \over \partial t}} & = & 0   & ; & 
 \bf{\nabla} \times \bf{B} - 
 \mu_0 \varepsilon_0 \ 
\displaystyle{{\partial \bf{E} \over \partial t}} & = & \mu_0
 \bf{j} \end{array}
$$

\medskip

\noindent Electromagnetic fields $\bf{E}$ and $\bf{B}$ are given by
these equations, once we give the distribution  of the sources $\rho \
\mbox{and} \ \bf{j}$, the electric charge and current densities
carried by the matter.  Since ${\bf E}$ and ${\bf B}$ are
defined through the force experienced by the charges and currents of the
matter, we can say that we have the complete description of the dynamics
of the electromagnetic process.  The underlying principle of
electromagnetism is  gauge invariance -- more precisely the local
gauge invariance.  $\bf{E}$ and ${\bf B}$ can be expressed in
terms of scalar $(\phi)$ and vector $(\bf{A})$ potentials, specified
together in the 4-vector notation as $A_\nu$.  The field $F_{\mu \nu}
(\equiv \partial_\mu A_\nu - \partial_\nu A_\mu)$ is invariant under the
gauge transformation that takes $A_\mu \rightarrow A_\mu - ie \partial_\mu
\Lambda (x)$.  Indeed the Maxwell's equations and all of physics is left
invariant by the gauge transformation.  

Symmetries in Physics, we recall are
closely related to conservation laws. The symmetry of gauge
invariance in electrodynamics is really the result of  the   
conservation of 
electric charge. While on the mundane level this implies that the sum of 
the charges remains the same during an interaction, on a more subtle 
plane, this calls for 
 an abstract operator $Q$, whose eigenvalue is indeed the 
charge of the state, and the gauge transformation is effected by the 
change of phase of the complex field that represents a charged particle:  
 $\psi(x) \rightarrow exp(iQ\Lambda) \psi$ is a gauge transformation on 
$\psi$, that is labelled by the parameter $ \Lambda$.    It is 
easily seen that 
successive transformation with parameters $\Lambda_1$ and $\Lambda_2$ will 
also 
be another gauge transformation with parameter $\Lambda_1 + \Lambda_2$ and 
the set of all gauge transformations will constitute  an abelian group. 
Since $\{e^{i\Lambda}\}$ is a set of all unitary unimoduar matrices of 
dimension unity, the relevant group is knows as  $U(1)$\footnote {$U(n)$ 
denotes a group consisting of unitary unimodular matrices of dimension $n$ 
and $SU(n)$ implies further that the matrices are of unit determinant.}. The 
charge operator $Q $ is said to be the generator of the group.

\medskip

When the symmetry is realised not only globally, but also at each space time 
point, -- that is when  the parameter is a space-time dependent 
$\Lambda(x)$ --  we have what used to be referred to as the gauge 
invariance of the second kind or more appropriately a local gauge symmetry.  
Roughly this means that the charge is  conserved at each space-time 
point. Naturally there is a need to propogate the information of this gauge 
transformation over space and time and this  necessitates the 
introduction of the 
gauge field $A_\mu (x)$; the electromagnetic interaction is indeed the
coupling of $A_{\mu}$ to the electric current density $j^\mu (x)$ of the
matter.  When quantised, $A_\mu$ represents the field for photon, the quantum
content of the electromagnetic waves as well as the messenger of the 
electromagnetic interaction.

\medskip

Standard model enlarges the gauge symmetry to include both weak
interactions and strong interactions.  First let us take into account
weak interactions and form Electro-weak gauge theory with the gauge group
as $SU(2)_L \times U(1)_Y$. This has an  $U(1)$ component, as was
encountered in the electrodynamics, but now related to another kind of
charge; call it weak hypercharge $Y_W$.In addition, we have the $SU(2)$ 
component of the symmetry group that characterises yet another local 
symmetry, which we may refer as the weak isospin\footnote {The 
weak isospin is different from the usual {\it isospin} which is believed 
to be the underlying symmetry responsible for the masses of proton and 
neutron to be nearly the same and also recognise them as two different {\it 
flavours} of Nucleon. Both weak isospin and (flavour) isospin have 
symmetry properties similar to the rotational symmetry that leads to the 
conservation of angular momentum.} ${\bf I}_W$. There are 
four generators of the electroweak symmetry, three for $SU(2)_L$ and one
for $Y_W$.  The elements of the symmetry group are
given as exp $i \Lambda_a I_W^a ; a = 1, 2, 3 \ \mbox{and exp} \ 
i \Lambda Y_W; I_W^a$ are like
the angular momentum operator $J^i, i = 1, 2, 3 \ \mbox{and} \ Y_W$ is like
the charge operator $Q_{em}$ ; $\Lambda_a \ \mbox{and} \ \Lambda $ are real
space-time dependent parameters. The symmetry $SU(2)_L \times U(1)_Y$ 
is thus given by four generators $I_W^a$ and $Y_W$; further the states 
are labelled by the irreducible representations of $|{\bf I}_W|^2, 
 I_W^3$ and $Y_W$, very much like the symmetry under rotations imply 
states with definite 
$ |\bf{J}|^2 $ and $ J^3 $ eigenvalues.  

\medskip

There is an important difference between  this  electroweak symmetry and 
the elecromagnetic $U(1)_{em}$ symmetry.The electroweak symmetry is said 
to be a hidden symmetry, 
meaning  that, while the dynamical Lagrangian is symmetric the solution to 
the equation of motion does not respect it entirely.  We say that the
symmetry is spontaneously broken down to $U(1)_{em}$ which is the
residual symmetry that represents 
the electric charge conservation and gives rise to
electrodynamics. To take an analogy, while roads admit traffic in either 
direction, we choose a convention , say in India, that the traffic moves 
along the {\it left} side of the road. Indeed it is equally possible to 
choose the opposite convention and have the traffic move along the {\it 
right} side, as in fact it does in several other countries. While the 
roads are symmetric, the `solution' we pick makes us lose it 
spontaneously. In the electroweak case, the symmetry $SU(2)_L \times 
U(1)_Y$ is {\it hidden} in the sense that the gauge fields exist 
corresponding to each of the four generators, $I_W^{1 \pm i2}, I_W^3 $ 
and $Y_W$, but only one of them (which corresponds to the residual 
conserved electric charge) survives as an unbroken symmetry. 

The 
generator of this symmetry $Q_{em}$ is a 
particular combination of the generators of the electroweak symmetry:

\begin{eqnarray}
Q_{em} \ = \ I_W^3 + {Y_W \over 2}
\end{eqnarray}

\noindent Thus photon field $(A_\mu)$ is a gauge boson, with a component from
$SU(2)_L$ gauge field $(W^3_\mu )$ and the $U(1)_Y \
\mbox{field} \ B_\mu$.

\medskip

\begin{eqnarray}
A_\mu \ = \ \sin \theta_W W^3_\mu \ + \cos \theta_W B_\mu 
\end{eqnarray}

\medskip

\noindent where $\theta_W$ is called Weinberg angle, a parameter to be
determined by experiment.  

The orthogonal combination $Z_\mu (= \cos
\theta_W W^3_\mu - \sin \theta_W B_\mu)$ as well as
$W^\pm_\mu ( = W^1_\mu \pm iW^2_\mu )$ 
will be different; for instance they acquire mass as a consequence of of 
the symmetry breaking. They give rise to 
 neutral
$(Z)$ and charged $(W^\pm)$ vector bosons, responsible for mediating weak 
interactions. The relative weak strength of the {\it weak processes} is 
due to heavy mass of these bosons that occurs as  the  the energy 
denominator in 
the relevant amplitudes.   
Standard Model predicts that
such mesons will have masses in the region of 80 GeV  and this was confirmed 
when CERN 
experiments were able to actually see them in a spectacular 
experiment in 1980. In the $Sp \bar{p}S$ collider, when protons
and antiprotons were made to collide with a total energy of 540 GeV, it
was enough to produce $W^+ W^-$ pairs.  Later at LEP (Large Electron
Positron collider) it was possible to have \ $e^+ \ \mbox{and} \ e^-$ collide to produce $Z$
meson, when the energy reached 91 GeV.  Thus we have the complete set of
messengers  of electro weak forces: Photon $(m_\gamma = 0)$, Charged Vector
Bosons $(m_W^\pm  = 80 GeV)$ and Neutral Vector boson $(m_Z = 91 GeV)$.
At energy scales large compared to 100 GeV, in the region when we may 
regard  $W$ and $Z$ mass to be small compared to their momenta, the 
strength  of electromagnetic and the `weak' interactions will be of
comparable order, thus resulting in the manifestation of the full
electroweak symmetry.

\medskip

The gauge bosons that are responsible for the electroweak forces couple to 
the electromagnetic, charge-changing and neutral weak currents, 
the coupling parameter (which is actually mildly distance  dependent)
of $W^a_\mu \ \mbox{and} \ B_\mu$ may be denoted as $g$ and $g^\prime$.  The
electromagnetic coupling $\displaystyle{\alpha \left( \equiv {e^2 \over 4\pi} \right) = 
{1 \over 137}}$ is related
to them through 

\medskip

\begin{eqnarray}
e \ = \ g \sin \theta_W = g^\prime \cos \theta_W
\end{eqnarray}

\medskip

\noindent From a variety of experimental inputs one finds that $\sin^2
\theta_W \simeq 0.23$.

\medskip

Now let us turn to the matter content of the universe.  Basic
building blocks of matter come in two types of spin $1 \over 2$ fermions --
leptons and quarks.  The most familiar leptons (which in Greek, 
means light particle) are electrons ($e^-$) and the associated neutrino
$\nu_e$ that is emitted with an electron in a $\beta$-decay (say, when
$n \rightarrow p + e^- + \bar{\nu}_e)$.  For a spin half state there are
quantum states with spin up and down (with respect to some definite axis,
say z-axis) or equivalently left and right handed helicity states. In 
discussing electroweak theory, it is convenient to group left helicity 
states and right helicity states seperately, since experiments show 
that the right helicity fermionic states do not take part in the weak 
interactions. The weak isopspin (${\bf I}_W$)  symmetry involves only 
left helicity fermions and hence is the subscipt L in referring to the 
weak isospin  $SU(2)_L$ symmetry. While the right helicity fermions 
such as  
$e^-_R$ is a $SU(2)_L$ singlet, left helicity fermions occur  as $SU(2)_L 
$ doublets. 
The neutrino $\nu_e$, which has left helicity and  the left 
handed $e^-_L$, form a  electroweak $SU(2)_L$
doublet, very much like the spin up and down states of angular momentum 
representaion. 
This means that $I_W^3$ quantum numbers for $\nu_e, e^-_L$ and $e^-_R$ are 
respectively $+{1 \over 2}$, $ -{1 \over 2}$ and $0$. 
Since $\displaystyle{Q_{em} = I_W^3 + {Y_W \over 2}}$, we find that the 
doublet $\left(
\begin{array}{c}\nu_e \\ e^-_L \end{array} \right)$ has weak hypercharge 
$Y_W$ = -1 and $e^-_R \ \mbox{has} \ Y_W = -2$.  That the left
and right handed fermions have different electroweak group
properties is at the root of the parity non-conservation of the weak
interaction.  In particular there is no right handed neutrino at all.
Nevertheless, notice that both $e^-_L \ \mbox{and} \ e^-_R$ have the
same value for $Q_{em}$, which fact ensures that the electromagnetic part 
has parity invariant coupling. 

\medskip

\medskip

Now turning to the strongly interacting states (collectively referred as
hadrons) such as protons and neutrons, the sub constituents are $u$ and $d$
quarks.  In a naive quark model proton is made up of two $u$-quarks and one
$d$-quark and neutron has two $d$-quarks and one $u$-quark.  Since nucleons
have an attribute of baryon number, quarks must have
nucleon or baryon number of ${1 \over 3}$ unit; further 
$u$ has $+{2 \over 3}$ units
of electric charge and $d$ has $-{1 \over 3}$ units.  As far as the
electroweak group properties are concerned, like  leptons, left handed
quarks are $SU(2)_L$ doublets and the right handed ones are neutral under
$SU(2)_L$ group, again reflecting the feature that only left handed 
fermions have a role in weak interaction.  We may 
read off their $Y_W$ quantum number from Eqn (2) and fill up the table 1.

\begin{table}[h]
$$
\begin{array}{|c|c|r|} \hline
\mbox{State} & I_W^3 & Y_W \\
             & \mbox{of} \ SU(2)_L & \\  \hline
 & & \\
\left(
\begin{array}{r}
u_L \\ d_L \end{array}\right) &
\left(
\begin{array}{r}
{1 \over 2} \\ -{1 \over 2} \end{array} \right) &
{1 \over 3} \\
u_R & 0 & {4 \over 3} \\
d_R & 0 & -{2 \over 3} \\
\left(
\begin{array}{r}
\nu_e \\ e^-_L \end{array} \right) &
\left(
\begin{array}{r}
{1 \over 2} \\ -{1 \over 2} \end{array} \right) &
-1 \\
e^-_R & 0 & -2 \\ \hline
\end{array}
$$
\caption{Quantum numbers of basic fermionic states}
\end{table}

\medskip

While protons and neutrons are three quark states, the mesons,
such as pions $(\pi^\pm, \pi^0$ of mass 139 and 135 MeV respectively) are
quark-antiquark bound states.  Indeed the plethora of resonant states
that were observed in the high energy collisions in the 50's and 60's can
{\it {all}} 
be classified as either ($qqq$) baryons or ($q\bar{q}$) mesons.  
Before proceeding further, we need to address 
an issue that arises since quarks are fermions and
hence should satisfy Pauli's exclusion principle.  Most prominent excited
state of nucleon is a spin $\left({3 \over 2}\right)$ and isospin
$\left({3 \over 2}\right)$ quartet $(\Delta^{++}, \Delta^+,
\Delta^0, \Delta^-)$ with mass $M_\Delta = 1238$ MeV and width
$\Gamma_\Delta = 120$ MeV.  This is, like nucleon, a three quark s-wave
bound state.  In particular $\Delta^{++}$ state with $m_s = {3 \over
2}$ will be represented by a symmetric wave function of three  $u$-quarks,
with all the spins aligned along the same direction.  Such a state cannot be
there, forbidden by Pauli's exclusion principle, except if the quarks
possess yet another quantum number, in the internal space of which the wave
function of the $\Delta$- state  
should be antisymmetric.  This space is referred to as the color\footnote 
{We deliberately use the American spelling to denote that this 
has nothing to do with the usual colour  that stands for the
variety of hues.} space and let the color index of quark $q_i$ take values
1, 2, and 3.  In order that $\Delta^{++}$ wave function is
antisymmetric under the interchange of all quantum numbers of constituent
quarks, it is necessary that it is of the form $\epsilon_{ijk} u_i u_j u_k$ in
the color space.

\medskip

This color symmetry can be promoted as a local gauge symmetry, with the
gauge group as $SU(3)$ and it is remarkable that the forces related to this 
symmetry are  the
basic dynamics underlying all {\it strong interactions} -- a further 
example of symmtry being the essence of dynamics.  Quarks are color
triplets and strangely they are {\it never} observed in its free form, 
and all known strongly interacting states like nucleons and 
mesons are   
color singlet bound states of quarks and antiquarks.  
  That is, the physical states that are
asymptotically realised are either  states that  do not
have any color substructure, such as electrons and neutrinos or hadrons
(eventhough the strong interaction properties of them are due to color 
forces of the components) that are color neutral states of the type 
${\displaystyle{\sum_{ijk} \epsilon^{ijk}  q_i q_j q_k}}$ \ and \ 
${\displaystyle{\sum_i q_i \bar{q}^i}}$.  The strong interactions between nucleons and mesons are the
color analogues of van der Waals forces that were seen responsible for the residual electromagnetic effects between charge neutral atoms and
molecules.  The role of photon is played in the color interaction  by an  
octet of colored massless vector bosons
that are known picturesquely as  gluons.  Being colored states gluons are 
also not observable as asymptotic physical states.

\medskip

\begin{table}[h]
$$
\begin{array}{|c|c|} \hline
\mbox{Vector Meson} & \mbox{Associated} \\ 
                    & \mbox{generators} \\ \hline
 & \\
\mbox{photon} {(\gamma )} & Q_{em}\\
W^\pm & I_W^{1 \pm i2} \ \mbox{of} \ SU(2)_L \\
Z & I_W^3, Y_W \\
\mbox{gluons} & T^a a = 1,\ldots 8 \ \mbox{of} \ SU(3)_C \\
\hline
\end{array}$$
\caption{Gauge Bosons}
\end{table}

In summary, the Standard Model is a  gauge field theory with
the local symmetry of $U(1)_Y \times SU(2)_L \times SU(3)_C$ spontaneously
broken down to $U(1)_{em} \times SU(3)_C$.  The twelve (1+3+8) generators
of the symmetry group are related to the 12 vector bosons (photon $(\gamma), 
W^\pm, Z$, octet of gluons) that mediate various interactions.  Of
these, photon and gluons are massless and represent the unbroken 
symmetries of
Quantum Electro Dynamics (QED) and Quantum Chromo Dynamics (QCD).  The
matter consists of a set of leptons $(e^-, \nu_e)$, color 
triplet quarks ($u_i,
d_i$) and the corresponding antiparticles.  Color confinement is believed
to be an important, still to be properly understood property of the $SU(3)$ 
gauge theory.

\vskip 0.3in

\noindent {\underline {MORE A-TOMS?}}

\vskip 0.2in

In the preceeding section, we have described the vector bosons as 
responsible for
all interactions and the main matter ingredients to be one set of quarks
and leptons.  As we probe shorter distances, through high energy
collisions, we find that there appears to be two more copies of the set of
fermions and they are referred to as the second and third generations of
matter.  First member of the second set to be discovered (during 50's in 
cosmic ray studies) was the muon
$(\mu)$, which is just a fat electron, about 200 times heavier and this was
seen 
accompanied by a nearly massless neutrino, call it $\nu_\mu$.  Like $u$ and
$d$ quarks of the electron family, in the second family, we have a pair of
quarks, which are denoted as $c$ and $s$, and they carry Charm and Strangeness
quantum numbers respectively.  Since the quarks are not realisable as
asymptotic states one can not talk about their mass; nevertheless we may
associate with $s$-quark an effective
 mass of about 160 MeV and $c$-quark about 1500 MeV,
gleaned from the analysis of deep inelastic 
scattering probes or mass content of the states carrying strangeness and
charm.  Apart from the mass parameters and the flavour\footnote {The 
flavour of the state is given by the type of the constituent quarks. The 
$u$ and $d$ quark indicate the flavour isospin doublet. Strangeness $S$ 
and Charm $C$ flavours are carried by $s$ and $d$ quarks.} quantum 
numbers $S,
C$ etc, all fermions of the 2nd generation have identical structure and
properties with respect to the symmetries of the Standard Model.  In
particular the left handed fermions are $SU(2)_L$ doublets and the right
handed fermions are singlets.  Muon, that has a life time of about
$10^{-6}$s, undergoes decay to electron and neutrinos; $\mu^-
\rightarrow  \nu_\mu + e^- + \bar{\nu}_e$, similar to the interactions $d
 \rightarrow u + e^- + \bar{\nu}_e$, which happens during the $\beta$- decay
of free neutron $(n \rightarrow p + e^- + \bar{\nu}_e)$.  
Strangeness carrying hadrons arise when we have a strange quark in
the state.  For example, some of the strangeness carrying 
hadrons are $\Lambda (1115 MeV)$, which
is a $(uds)$ bound state and $K^{+,0}$ (495 MeV), which are $(u \bar{s})$
and $(d \bar{s})$ mesons. From the fact that a non-strange vector meson
$\phi(1020 MeV)$ decays almost entirely into the channels $K^+ K^- \
\mbox{and} \ K^0 \bar{K}^0$, we conclude 
that it is mostly a $(s \bar{s})$ state.
  In 1974,  the new flavour charm was discovered, when the 
 charmonium state ($c \bar{c}$) was
observed as a very narrow resonance at about 3097 MeV in the $(e^+ e^-)$
collission.  Subsequently charmed baryon $\Lambda_c (2281)$ MeV as a 
$(udc)$ bound state, $D^{\pm,0}$ (1869 MeV) mesons of ($c \bar{d}$) and $(c
\bar{u})$ states etc. were discovered.

\medskip

The Third set of fermions starts with the $\tau$- lepton at 1760 MeV and
the related neutrino $\nu_\tau$ in the lepton sector.  In the quark part, in
the late seventies, a new flavour B was found as a consequence of yet
another quark, call it $b$-quark.  We may call it `beauty' or `bottom' (to
denote that it is a quark with charge $-{1 \over 3}$ and occurs as the
bottom part of the $SU(2)_L$ doublet).  Since Beautionium is a
$(b\bar{b})$ state with 9650 MeV, it is natural that `beautiful' states
will have masses in the range of 5 GeV.  $B^{+,0}$ mesons (5275 GeV) are
$(u\bar{b})$ \ and \ $(d\bar{b})$ states, while $\Lambda_b$ (4750 MeV)
is a $(udb)$ bound state.

\medskip

The third set needs another quark to complete the picture.  Such a quark
will occupy the top part of the $SU(2)_L$ doublet, of which the $b$-quark is the
bottom.  The quest for the top quark $t$ (which is also referred to
as truth,  the related flavour) had been elusive, mainly because the top
quark turns out to be much heavier than was expected.  Whereas $u,d,s$ are 
relatively light
quarks $c$ and $b$ were found to be 1.5 and 5 times as heavy as nucleon 
(whose mass is about 1GeV).  It
was not possible to sight $t$-quark until we increased the collision
energies considerably; the Fermilab results last year suggest that the
$t$-quark is about 175 GeV.  With the discovery of
$t$-quark the third set of fundamental fermionic matter appears complete.
Notwithstanding the much heavier masses, the third set of leptons $\nu_\tau \
\mbox{and} \ \tau$ and the third set of quarks $(t, b)$ have identical
structure and couplings with respect to the Standard Model interactions as
the first and second generations of fermions.

\vskip 0.2in

\begin{table}[h]
$$
\begin{array}{|cc|c|c|c|c|} \hline
\multicolumn{2}{|c|}{\mbox{BOSONS}} &  &
\multicolumn{3}{c|}{\mbox{FERMIONS}} \\
\hline
& & & \multicolumn{3}{|c|}{LEPTONS} \\ \cline{4-6}
Charge & & Charge & &  & \\
\pm 1 & W^{\pm} & 0 & \nu_e & \nu_\mu & \nu_\tau \\
      & (80.2 GeV) &   & (0) & (0) & (0) \\
& & & & & \\
0 & Z & -1 & e & \mu & \tau \\
  & (91.2 GeV) &  & (0.511 MeV) & (105.7 MeV) & (1.760 GeV) \\
& & & & & \\ \cline{4-6} 
& & & \multicolumn{3}{|c|}{QUARKS} \\ \cline{4-6} 
0 & Photon & 2/3 & u & c & t \\
  & (0) & & (5 MeV) & (1.500 GeV) & (175 GeV) \\
  & & & & & \\
0 & Gluons & -1/3 & d & s & b  \\
  & (0) & & (8 MeV) & (160 MeV) & (4.250 GeV) \\ \hline
\end{array}
$$
\caption{The a-toms of the Standard Model}
\end{table}

\medskip

In summary, we can now assert that, the a-toms of Democritus
 are the three sets of fermions; each
set consists of a left helicity lepton doublet $\left( \begin{array}{c}
\nu_E \\ E^-_L \end{array} \right)$, right helicity lepton $E^-_R \
\mbox{and} \ a$ color triplet of left handed quark doublet $\left(
\begin{array}{c}U_L \\ D_L \end{array} \right)$ and right handed
singlets $U_R \ \mbox{and} \ D_R$ with $E, U \ \mbox{and} \ D$ \ as the
generic labels.  All physical states are made up of
these fundamental sets of fermions and their dynamics is given by the 
Lagrangian that displays the gauge dogma.  The forces of interactions
are due to the messenger vectors bosons ($\gamma, W^\pm, Z$ and \ gluons)
which are also a-toms.

\vskip  0.3in

\noindent {\underline {\bf EPILOGUE}}

\vskip 0.2in

Can  there be more than three families of matter?  Why are there three
generations?  Can we understand the reason why top quark is much heavier
than the rest.  How many parameters do we need to specify in describing
the Standard Model?  Why should the gauge group be $U(1) \times SU(2)
\times SU(3)$?  Is there atleast a partial answer to all these queries?

\medskip

Strictly speaking in our description of the Standard Model, we are yet to
introduce mass parameters of the constituents.  The gauge fields, like
photons, to begin with  are massless.  
The symmetry does not admit a mass term for the 
chiral fermions\footnote {The term chiral fermion implies that the left 
and right helicity states of the fermion have {\it different} 
properties.} in the Lagrangian.  Indeed, if the gauge
symmetry is exact, there is no scope for any of the masses. Then how are
the masses generated.  Recall that
we mentioned that the $U(1) \times SU(2) \times SU(3)$ is spontaneously
broken down to $U(1)_{em} \times SU(3)_c$.  The mechanism that achieves
this should be the key to understanding the various mass values.  While the 
full Lagrangian is manifestly $U(1)
\times SU(2) \times SU(3)$ symmetric, the solutions (that includes the
ground state vacuum) has less symmetry.  For this purpose, we need one
more species of particles, which will be responsible for the symmetry
breaking and 
has an added role to make many particles massive.  The agent for
this is  a $SU(2)_L$ doublet
scalar field.  The Higgs doublet 
$
\Phi = {1 \over \sqrt{2}} \ \left(
\begin{array}{c}
 \phi_1 + i\phi_2 \\
 \phi_3 + i\phi_4 \end{array} \right)$
 has $Y_W = 1$. The Higgs self interaction is such that it
 has a nonvanishing vacuum expectation value $< \Phi > =
\left(
\begin{array}{c}
0 \\ v/\sqrt{2} \end{array} \right)$, where $v$ is an important
real parameter, the non zero value of which indicates 
the spontaneous breaking of the symmetry. To understand this let us 
consider as example, a single complex field $\phi$ and let the 
interaction be given by $V(\phi)$. Let $V(\phi) = {1 \over 2} 
m^2\phi^*\phi + {\lambda \over 4} (\phi^*\phi)^2$. If, as usual, $m^2 > 
0$ and $\lambda > 0$, this represents scalar meson of mass $m$ and 
quartic interaction of strength $\lambda$. The ground state will have $\phi 
=0$ and is unique. If, on the other hand $m^2 < 0$, then $\phi = 0 $ is 
no longer a minimum of $V(\phi)$. The minima, on the contrary are 
given 
by $|\phi| = (-m^2/\lambda)^{1/2} = v $. The ground state solutions for 
$\phi$ could assume values $v e^{i\theta}$; $\theta$ arbitrary. Hence the 
ground state is degenerate. While the full symmetry is reflected in the 
complete set of the solutions, as soon as we pick a particular solution, 
say $\phi = v$, then the symmetry, in this case $U(1)$, is spontaneously 
broken. In the Standard Model, similar process occurs and we have the 
solution for $\Phi$ reflect the symmetry breaking from $SU(2)_L \times 
U(1)_Y$ to $U(1)_{em}$. From $\mu$-decay rates, we determine 
that $v \simeq 250 GeV$. It is possible to recast so that, three 
of the four real scalar fields of $\Phi$ play the role
of the longitudinal polarisation modes of $W^+, W^- \ \mbox{and} \ Z$
bosons and in this way make these gauge bosons massive.  (Recall
photon  has only two transverse
polarisation modes while the massive vector mesons should have longitudinal
polarisation as well.)  The last remaining field of $\Phi$ 
should show up as a massive
scalar meson, called Higgs, that remains to be discovered.
Higgs mechanism also lets the fermions acquire mass through the coupling of 
the Higgs doublet to fermions $(\bar{\psi}_L \ \Phi \ \psi_R)$.
Closely related to the
fermion masses is the fermion mixing angles, three in number and the
possibility of intrinsic CP violating phase angle\footnote {C 
and P here denote two 
discrete symmetries - Charge conjugation that interchages particles and 
anti-particles and Parity that reflects the spatial co-ordinates. They 
were  believed to be good symmetries until they were both seen violated 
in the weak interaction, leaving  however the combination CP intact, 
except in 
very few processes involving the neutral K mesons. The phase angle is a 
measure of the CP violation.} in the mass matrix.  

\medskip

Standard Model thus
consists of a gauge symmetry with 12 gauge vector bosons, of which all but
three $(W^\pm, Z)$ are massless, three sets of quarks and leptons and one
neutral Higgs Scalar meson, that is yet to be discovered.  It is specified by
three gauge coupling parameters ($\alpha_{em}, \alpha_s \ \mbox{and} \
\sin^2 \theta_W$) and Higgs meson vacuum expectation value that
characterises the masses of $W^\pm \ \mbox{and} \ Z$, Higgs meson coupling
to fermions that could be parametrised in terms of the masses of the 3
charged leptons and  6 quarks, 3 mixing angles, one CP violating phase
and one yet to determined Higgs meson mass.  Thus we count in all merely
18 parameters that are needed to define the Standard Model.

\medskip

Perhaps there are no more than three generations.  $Z$ boson width, which
is a measure of the sum total of probabilities of $Z$ decays into various
modes, is known so well that it can accommodate no more than three species
of neutrinos.  Also there will be problems with cosmology if
there are more than four types of neutrinos.  If there were only two
families, there is no room for the CP violating phase -
whose presence was indicated in the 1965 experiment that found that there
is a tiny amount of CP violation in the neutral $K$- meson decays.  Further the
fact that the universe consists of only matter and no antimatter can be
triggered only if we may have the possibility of  
an intrinsic CP violation.  Are these then,
the reasons why we have three generations of matter?  

\medskip

We may have to go
beyond the Standard Model to get any clues about why this speicifc group
is chosen or why top quark, is so heavy.  Is there a hint in the fact that
the top quark, has mass in the same region as weak interaction symmetry
breaking scale?  Then, could it be that Higgs scalar (which is the agent
for breaking symmetry), is not an a-tom, but a ($t\bar{t}$) bound state?
We have to wait for about a decade, when the Large Hadron Collider will be
completed at CERN, and hopefully produce Higgs mesons copiously.  Until 
then, we
have just about three sets of fermions, and the twelve vector bosons 
of the gauge interaction 
as the a-toms that make up all of the universe and let the future
decide whether the Higgs Scalar boson is elementary or composite.

\vskip 0.3in

\noindent {\bf For further reading:}

\vskip 0.2in

\begin{enumerate}
\item Dreams of the final theory - Steven  Weinberg (Pantheon Books,
New York 1992).  A popular account at the Standard Model and journey
towards a final theory.

\item The God Particle - Leon B. Lederman (Dell Paperback 1995) \\
The Higgs Scalar meson is referred to by Lederman as the God
Particle and the popular book is on the history of developments in
Particle Physics and the expectation for the discovery of Higgs in the
first decade of the 21st century.

\item Building up the Standard Gauge Model of the High Energy Physics -
G. Rajasekaran in Gravity, Gauge theories and the Early Universe (UGC
Instructional Conference) Ed. B.R.Iyer {\it et al} (Kluwer Academic
Publishers 1989) p.185-236. \\
Lectures on the Standard Model and beyond.

\item  High Energy Physics - Special Section, Current Science {\bf 71}
(1996) p.109-127.  \\
Proceedings of a symposium on High Energy Physics in
21st Century held as a part of the 61st Annual Meeting of the Indian
Academy of Science during 10-12 November 1995 at Madras.  Articles are
by G. Rajasekaran, D. P. Roy, Romesh K. Kaul, Abhijit Sen and R.
Ramachandran.
\end{enumerate}
\end{document}